\documentclass[aps,pra,showpacs,floatfix,superscriptaddress,twocolumn]{revtex4-1}
\usepackage{graphicx}
\usepackage{dcolumn}
\usepackage{amsmath}
\usepackage{amssymb}
\usepackage{dsfont}
\usepackage{bm}

\begin{document}

\title{Entropy rate of message sources driven by quantum walks}

\author{B. Koll\'ar}
\affiliation{Wigner RCP, SZFKI, Konkoly-Thege Mikl\'os \'ut 29-33, H-1121 Budapest, Hungary}
\affiliation{Institute of Physics, University of P\'ecs, Ifj\'us\'ag \'utja 6, H-7624 P\'ecs, Hungary}

\author{M. Koniorczyk}
\affiliation{Institute of Mathematics and Informatics, 
University of P\'ecs, Ifj\'us\'ag \'utja 6, H-7624 P\'ecs, Hungary}

\pacs{03.67.Ac, 89.70.Cf, 05.40.Fb}

\date{\today}

\begin{abstract}
  The amount of information generated by a discrete time stochastic
  processes in a single step can be quantified by the entropy rate. We
  investigate the differences between two discrete time walk models,
  the discrete time quantum walk and the classical random walk in
  terms of entropy rate.  We develop analytical methods to calculate
  and approximate it.  This allows us to draw conclusions about the
  differences between classical stochastic and quantum
  processes in terms of the classical information theory.
\end{abstract}

\maketitle

\section{Introduction}

Source coding, that is, the encoding of the output of an information
source, is one of the fundamental problems of information theory. A
source emits a sequence of possible messages. A simple model of a
source thus consists of a discrete infinite sequence $\mathbf{X}$ of
random variables $X_k$, whose actual values $x_k\in \mathcal{X}$
describe the $k$-th message, whereas $\mathcal{X}$ stands for the
range of the random variables. (We shall refer to the steps of this
sequence as \emph{iteration steps} throughout this paper.) This is
termed as a stochastic process. Information theory provides asymptotic
lower bounds on the number bits required per message to encode the
output of such a source. These bounds are based on arguments involving
the asymptotic equipartition property, a consequence of the weak law
of large numbers. If the sequence $\mathbf{X}$ consists of independent,
identically distributed random variables $X$, the required
asymptotical number of bits per symbol is $H(X)$, where $H$ stands for
the Shannon entropy. If, however, the source is described by a generic
stochastic process, in the arguments regarding the above-mentioned lower
bound, the entropy is replaced by the so-called entropy-rate:
\begin{equation}
  \label{eq:entrate}
  H(\mathbf{X})=\lim_{N\to \infty}\frac1N H(X_1,X_2,\ldots X_N)\,,
\end{equation}
if the limit exists. It can be shown that for stationary time-invariant
stochastic processes, the entropy rate exists and is equal to a
similar quantity defined via conditional entropies
\begin{equation}
  \label{entropy_rate_def2}
  H'(\mathbf{X})=\lim_{N\to \infty} H(X_N|X_{N-1},X_{N-2},\ldots X_1)\,.
\end{equation}
Moreover, for a stationary time-independent Markov-chain it simplifies
to
\begin{equation}
 H(\mathbf{X}) =  \sum_{x_1\in \mathcal{X}} p(X_1 = x_1) H(X_2 | X_1 = x_1)\,.
\label{entropy_stationary_conditional}
\end{equation}
Expressed in terms of the probability transition matrix $P_{i\to j}$
where the indices $i,j$ label the elements of the range $\mathcal{X}$,
and the stationary distribution $\mu$ (that is, $\mu P=\mu$) of the Markov
chain, the entropy rate can be calculated as
\begin{equation}
  \label{entropy_stationary_markov}
  H(\mathbf{X}) = - \sum_{i,j} \mu_i P_{i\to j} \log_2 P_{i\to j}\,
\end{equation}
All such Markov
processes are equivalent to a classical random walk on an undirected
weighted graph, in which the probability transition matrix is
expressed from the weights $W_{i,j}$  as
\begin{equation}
  \label{eq:Pwgraph}
  P_{i\to j}=\frac{W{i,j}}{\sum_{j}W_{i,j}}\,.
\end{equation}
These facts, the details of which can be found in many
textbooks of information theory (e.g., Ref.~\cite{Cover2006}) motivate
our present investigation.

Quantum information theory is a nontrivial generalization of the
classical one; hence, one expects that the above arguments can be also
generalized that way. Indeed, there are various approaches of quantum
information to form the concept of quantum entropy rate
\cite{Alicki2001,Benatti2006,Benatti2007,Crutchfield2008}.  In the
present paper, however, we will focus on a description in terms of
classical information theory. We consider the following simple scenario.  Let us
assume that we have a source of information in a ``black box". We
know that there is some physical process inside, generating classical
messages. However, this process might be either a classical random
walk or some quantum process which generates classical information,
and has a well defined classical counterpart: If decoherence is
significant, it becomes a classical random walk. Then we can utilize
the above described apparatus of classical information theory to
compare the classical random walk with one of its quantum
generalizations. Hence, we do not go beyond the concepts of classical
information theory here; instead, we utilize them in order to learn
more about the classical-quantum transition: What is the difference
between a classical and quantum black box from the point of view of
entropy rates?

If one seeks a quantum process with a classical counterpart which is a
discrete-time random walk, the discrete-time quantum walk (QW) is a
suitable choice. Quantum walks \cite{Aharonov1993,Kempe2003} are nontrivial
generalizations of the classical random walks, obeying unitary, and
therefore deterministic, dynamics. This simple quantum model allows
researchers to study various physical phenomena, including transport
\cite{Mulken2007,Mulken2011,Anishchenko2012}, percolation \cite{Leung2010,Kollar2012,Darazs2013}, and
topological effects \cite{Kitagawa2010,Asboth2012,Asboth2013,Rudner2013}. Similarly to
classical walks, QWs can be naturally utilized for algorithmic
applications. The universality of QWs is proven in
Refs.~\cite{Childs2009,Lovett2010}. For an overview of quantum
informational applications of QWs, see Ref. \cite{Venegas2012}. In the
recent years, several experimental breakthroughs have been achieved
\cite{Karski2009,Schmitz2009,OBrien2010,Zahringer2010,Schreiber2010,Broome2010,Schreiber2012,Silberberg2012,Torres2012,Thompson2013,Alberti2013},
these experimental successes naturally motivate the theoretical study
of QWs.

This paper is organized as follows: First, we calculate the entropy
rate of certain classical random walks. Then we define quantum walks
and a scenario in which they serve as signal sources. We give explicit
and approximate methods to calculate the upper bound of entropy rate
and the actual entropy rate of the source considered. Finally, classical and
quantum cases are compared and conclusions are drawn.

\section{Entropy rate of some classical random walks}

Let us summarize the properties of certain classical walks, which we
shall refer to later in this paper.  The results presented here can be
obtained by a straightforward application of the definitions given in
the previous section.  

We remark here that for a general classical walk as a stationary
Markov process, the entropy rate, according to
Eq.~\eqref{entropy_stationary_markov}, is the average of the entropy
of the rows of the probability transition matrix taken with the
stationary probability of each vertex. (As each row corresponds to a
vertex where the walker may stand in a step, and each column to an
edge pointing to a possible vertex he can jump to.) In particular, if
the stationary distribution is uniform and, for some symmetry reason,
the rows are permutations of each other (thus having the same
entropy), the entropy rate is simply the entropy of a row. That is, in
the graph picture, the process is equivalent to a sequence of
independent identically distributed random variables describing the
random decision taken by the the walker in each step. This reasoning
is applicable in some of the cases we discuss here. An unbiased
(isotropic) classical random walk (CW) on a $d$-regular simple graph,
for instance, has the entropy rate of $\log_2 d$: wherever we find the
walker, it has $d$ equal-probability edges to follow (isotropy), and the
stationary distribution is obviously uniform. Hence, in this model,
for every step we need $\log_2 d$ classical bits to encode the
direction where the walker has moved randomly.

Now let us consider a one-dimensional walk (on a finite cycle with $M$
vertices) and suppose that we intend to encode the position only at
every $w$-th step of the walk. This reads
\begin{eqnarray}
H_w^{\text{CW}} & = & 2^{-w} \sum_{i = 0}^{w} \binom{w}{i} \left\{ w - \log_2 \binom{w}{i} \right\} \nonumber\\
& \approx & \frac{1}{2} \left( -1 + \log_2 \pi e w \right) \,,
\label{entropy_classical}
\end{eqnarray}
which is the Shannon entropy of the binomial distribution and the
Gaussian distribution respectively.  We derive formula in
Eq.~\eqref{entropy_classical} later, concluding in
Eq.~\eqref{delta_erate}.  Note, that the $1/2$ prefactor is a
consequence of the diffusive spreading of the CW.  The parameter $w$
will be termed as ``waiting time'' in what follows, which will also be
the time we wait between two subsequent quantum measurements in the
corresponding quantum protocol.  Note that
Eq.~\eqref{entropy_classical} is valid for both infinite and finite
systems, as long as $w\ll M$. For finite $M$ and that are rates high enough (in
one-dimensional (1D) cycle graphs, for instance, this occurs for $w > M/2$), the walker
mixes with itself, making the rate given by
Eq.~\eqref{entropy_classical} inaccurate. In this case the sequence
becomes a series of independent random variables with a uniform
distribution over the accessible positions, thus the entropy rate
becomes the upper bound of the possible entropy rates,
\begin{equation}
H_{\textrm{limit}} = \left\{
 \begin{array}{cc}
 \log_2 M &\text{for}\, \text{odd} \, M  \\
 -1 + \log_2 M &\text{for}\, \text{even} \, M  
 \end{array}
 \right. \,.
\label{entropy_limit}
\end{equation}
The difference caused by the parity is due to the fact that the
positions accessible for the walker may be restricted. In a
1D cycle graph with even number of sites ($M$), the walker, from a
given position, can reach either the even or the odd labeled sites
only, depending on the waiting time $w$. Therefore, even for the
limiting $w \gg M$, only half of the sites can be reached by the
walker. For cycles with an odd number of sites, this restriction does
not hold.

The system under consideration is translationally invariant
(homogeneous in space): The transfer probabilities $P_{x \rightarrow x +
  \delta}$ between arbitrary lattice sites $x$ and $x + \delta$ depend
only on the difference (distance) $\delta$ of the two sites.
Thus, we introduce the probability of a $\delta$ length shift 
\begin{equation}
p(\delta) \equiv P_{x \rightarrow x + \delta}\,.
\label{delta_equivalence}
\end{equation}
In systems obeying this symmetry, it is common to encode the
difference $\delta$ of the actual random position outcome from the
previous random outcome, leading to the usage of at most $w+1$
symbols, thus a finite alphabet.  It is straightforward to see
that the two encoding methods --- encoding the absolute position
outcomes and encoding the relative position differences --- are
equivalent.  From (\ref{entropy_stationary_markov}) and
(\ref{delta_equivalence}) one can readily calculate the entropy rate
as
\begin{equation}
H^{\text{CW}}_w = - \sum_{\delta=-w}^{w} p(\delta) \log_2 p(\delta)\,,
\label{delta_erate}
\end{equation}
which after a short calculation results in
Eq.~(\ref{entropy_classical}).  We can conclude that the entropy rate
of a process arising from a one-dimensional classical walk with
waiting time $w$ is simply the Shannon entropy of the distribution of the shifts.
Note that for the sake of readability the sum in
Eq.~\eqref{delta_erate} is taken between $-w$ and $w$; however, since
the classical walker leaves its position in every step, there is a
parity correspondence between $w$ and $p(\delta)$, thus we have $w+1$
symbols to encode at most.  In the next section we extend the concept
of entropy rate to sources driven by quantum walks by following the
procedure presented above.

\section{Discrete time quantum walk as a source of messages}

Given a $G(V,E)$ $d$-regular simple graph, the Hilbert space of a discrete-time coined quantum walk (QW) is defined as
\begin{equation}
\mathcal{H} = \mathcal{H}_P \otimes \mathcal{H}_C\,,
\end{equation}
where $\mathcal{H}_P$ is the position space corresponding to the vertices of the graph and $\mathcal{H}_C$ is the coin-space corresponding to an internal, ``coin" degree of freedom, i. e., directions pointing to the nearest neighbors.
Let us use the following shorthand for Hilbert-space vectors:
\begin{equation}
| v \rangle_P \otimes | c \rangle_C \equiv | v,c \rangle\,.
\end{equation}
The discrete-time unitary evolution is given by
\begin{equation}
 U=S \cdot \left( I_P \otimes C \right)\,,
\end{equation}
where 
\begin{equation}
 S = \sum_{v \in V,\, c \in [1..d]} | v \oplus  c, c\rangle \langle v, c |\,.
\end{equation}
$I_P$ stands for the identity operator on the position space, and $C \in SU(d)$ is the coin
operator acting on the internal degree of freedom. The abstract sum
$v \oplus c$ represents the nearest neighbor of the vertex
$v$ in the direction indicated by the coin state $c$.

We wish to use this deterministic quantum process as the source of
messages (classical random variables). Thus, we
introduce measurement into the system, closely following the procedure
we employed for the classical case: We let the walker evolve for $w$
steps and we measure its position afterwards. Should someone measure
the position of the walker, she will get a random position
$x$ with probability
\begin{equation}
p(X_k = x ) \equiv  \sum_{c'} \left| \langle x, c' | \psi_k \rangle \right|^2 \,,
\label{stochprocc_def}
\end{equation}
where  $| \psi_k \rangle = U^w | \psi_{k-1} \rangle$ is the Hilbert vector corresponding to the quantum state of the QW at the $k$th iteration step.
The corresponding $X_k$ is the random variable describing the position outcome at the $k$th iteration.
From now on, we consider the sequence of $X_k$ random variables as the stochastic process generating the message we wish to encode efficiently.
A similar model is considered in Ref. \cite{Shikano2010}, where the authors address the randomness induced by the frequent measurements. However, our case is differs fundamentally as we do not reset the coin state after every measurements and we focus on the entropic properties of the system.

Like classical walks, the QWs considered in the present paper are translationally invariant:
\begin{equation}
\langle y, c | U^t | x, c \rangle \equiv \langle y \oplus \delta, c | U^t | x \oplus \delta, c \rangle\,,
\label{homogeneousity}
\end{equation}
for all $x, y,t,\delta$ and $c$-s.  Consequently, in place of
encoding the $x_k$ measurement outcomes, one can encode position
differences $\delta = x_k - x_{k-1}$. Note, that
this encoding simplification does not affect the value of the entropy
rates, it is just the standard notation for systems with translation invariance. Equivalently, the original
problem can be rephrased so the black box outputs the relative
position differences $\delta$ instead of absolute positions.

The proposed definition of a QW-driven message source has a
well-defined classical connection: Should one consider an unbiased coin
matrix $C$ (with all complex elements having the same absolute value
in the natural (computational) basis), a quantum walk measured in
every single step (waiting time $w = 1$) behaves exactly like a
classical unbiased (isotropic) walk.

Throughout this paper we will investigate 1D QWs and use the $2 \times 2$ Hadamard matrix as the coin operator,
\begin{equation}
C_H = 
\frac{1}{\sqrt{2}} \left( \begin{array}{rr}
1 & -1 \\
1 & 1
\end{array} \right)\,,
\label{coin_hadamard}
\end{equation}
in most of the cases, unless stated otherwise.
We use the Hadamard coin since it is unbiased, thus we have a very well controlled quantum-classical transition at our hands. 

\section{Entropy rate of 1D quantum walks}
\label{precapproachsection}
Let us calculate the entropy rate of the process defined in the previous section. We note, that here we address the 1D problem, but the methods we give could be generalized for higher dimensional QWs.
First, we show a way to calculate the joint probability distribution $p(x_N, x_{N-1}, \ldots , x_1)$ of the possible outcomes. 
Employing Eq. (\ref{stochprocc_def}), the joint probability distribution of the random variable sequence $X_k$ is given by
\begin{widetext}
\begin{eqnarray}
\label{joint_distrib}
 p(x_N, x_{N-1}, \ldots , x_1) & = & \mathrm{Tr} \left( S_{x_N} U^w  S_{x_{N-1}} U^w  \ldots S_{x_1} U^w \rho_0 (U^w)^{\dagger} S_{x_1}   \ldots (U^w)^{\dagger} S_{x_{N-1}}  (U^w)^{\dagger} S_{x_N}  \right)\,,
\end{eqnarray}
\end{widetext}
where
\begin{equation}
S_{x} = | x, R \rangle \langle x, R | + | x, L \rangle \langle x, L |
\label{position_projector}
\end{equation}
is the projector of the von Neumann measurement corresponding to the position $| x \rangle_P$ and
$\rho_0 = | 0, c_{0} \rangle \langle 0, c_{0} | $ is the initial state of the 1D QW in the black box.
Next we employ the definition in Eq.~(\ref{eq:entrate}) to obtain the
entropy rate. Note, that we have to use the original definition as we are not
considering a Markovian process here.

However, calculating (\ref{joint_distrib}) and therefore the actual
entropy rate in the asymptotic limit is quite demanding both
numerically and analytically. In the following we present a method to
make the calculation manageable. It is based on the fact that the
transition probabilities between subsequent measurement outcomes
depend on a parameter which in fact can be taken into account. It is
the internal quantum coin state, which carries additional information in the following sense.

After every position measurement, the wave function collapses to a
single position site, but the information carried in the coin degree
of freedom that particular site survives the process: It serves as the
initial coin state in the following iteration. After acquiring any
position measurement outcome (a black box output) $X_k = x$,
since the evolution of QW is unitary (deterministic) until the position
measurement, the full quantum state of the actual collapsed QW can be
reconstructed with the knowledge of the full quantum state of the
preceding (initial) iteration. In summary, the coin degree of freedom
serves as a memory, carrying some information about the previous
steps. The importance of this observation is twofold: First, the
information carried in this internal memory can be used to improve our
encoding method. Second, we use the coin to aid our calculation of the
joint probability distribution, thus the entropy rate.

To employ the coin as a hidden continuous parameter of the model, we
introduce an extended, $P_{x, \alpha \rightarrow y}$ stochastic
transition matrix, where $\alpha$ is an abstract continuous parameter
representing an internal coin state. The definition is given as
\begin{equation}
\label{matrix_generalized}
P_{x, \alpha \rightarrow y} \equiv \mathrm{Tr} \left\{ S_y  U^w | x , \alpha \rangle\langle x , \alpha | (U^w)^{\dagger} \right\} \,.
\end{equation}
Since we know the initial (previous) quantum state of the system, the quantum state of the next iteration step can be calculated as follows: 
\begin{equation}
 | y \rangle_P \otimes |  \mathcal{C}(x,y,\alpha) \rangle_C \equiv \frac{S_y  U^w | x , \alpha \rangle}{\sqrt{\mathrm{Tr} \left\{ S_y  U^w | x , \alpha \rangle\langle x , \alpha | (U^w)^{\dagger} \right\}}}\,,
\end{equation}
where we defined function $C(x,y,\alpha)$ giving the unambiguous coin state.
Employing these definitions while using Eqs. (\ref{joint_distrib}) and (\ref{position_projector}) we arrive at
\begin{eqnarray}
p(x_N, x_{N-1}, \ldots , x_1) =\nonumber\\
P_{0,c_0 \rightarrow x_1} P_{x_1, c_1 \rightarrow x_2} P_{x_2, c_2 \rightarrow x_3} \ldots P_{x_{N-1},c_{N-1} \rightarrow x_N}\,,
\nonumber\\
\end{eqnarray}
where $c_i = \mathcal{C}(x_{i-1},x_i,c_{i-1})$ and $c_0$ corresponds to the initial coin state.

Let us use the translation invariance (\ref{homogeneousity}) of the system. We shall see that
\begin{equation}
p_c(\delta) \equiv P_{x,c \rightarrow x+\delta} =  P_{y,c \rightarrow y+\delta}
\label{pcdelta_def}
\end{equation}
and
\begin{equation}
\mathcal{C}(\delta,c) \equiv \mathcal{C}(x,x+\delta,c) = \mathcal{C}(y,y+\delta,c)
\end{equation}
for all values of $x$, $y,$ and $\delta$.
Thus
\begin{eqnarray}
p(x_N, x_{N-1}, \ldots , x_1)
=p(\sum_{i=1}^N \delta_i, \sum_{i=1}^{N-1} \delta_i, \ldots , \delta_1) \nonumber\\
=p_{c_0}(\delta_1) p_{c_1}(\delta_2) \ldots  p_{c_{N-1}}(\delta_N)\,,
\nonumber\\
\end{eqnarray}
where $c_i = \mathcal{C}(\delta_{i},c_{i-1})$ and $\delta_i = x_i - x_{i-1}$ with $x_0 = 0$.
Note that the product form of the probability  shows the true Markov chain like nature of the system: The probability of any outcome can only depend on the previous quantum state of the system, that is, the internal coin state and its position (which is in the $\delta$ difference picture is neglected due to translation invariance).
The Shannon entropy of the joint distribution can be calculated using the chain rule as 
\begin{eqnarray}
H (X_N, X_{N-1}, \ldots , X_1) 
=  \sum_{i=1}^N H (X_i | X_{i-1}, \ldots , X_1) \nonumber\\
= - \sum_{i=1}^N  \sum_{\alpha \in \text{CS}} \nu_{i-1}(\alpha) \sum_{\delta=-w}^{w} p_{\alpha}(\delta) \log_2 p_{\alpha}(\delta)\,,
\nonumber\\
\end{eqnarray}
where 
\begin{equation}
\nu_i(\alpha) = \sum_{\delta=-w}^{w} \sum_{\beta \in \mathds{C}(\delta,\alpha)} \nu_{i-1}(\beta) p_{\beta} (\delta)\,,
\label{coindistrib}
\end{equation} is the distribution of coin states at the $i$th iteration step,
with $\mathds{C}(\delta,\alpha) = \left\{ \beta \in \text{CS} \,|\, \mathcal{C}(\delta,\beta) = \alpha \right\}$ and $\nu_0 (\alpha) \equiv \delta_{\alpha, c_{0}}$.
In our notation, the $\delta$ symbol with two indices ( $\delta_{\alpha,c_0}$) is the Kronecker $\delta$.
By $\text{CS}$ we denote the continuous set of all abstract coin states.
The entropy rate is then given by taking the limit as in Eq.~(\ref{eq:entrate}),
\begin{eqnarray}
H(\mathbf{X})=\lim_{N \rightarrow \infty} \frac{1}{N} H (X_N, X_{N-1}, \ldots , X_1)  \nonumber\\
= - \lim_{N \rightarrow \infty} \frac{1}{N}  \sum_{i=1}^N  \sum_{\alpha \in \text{CS}} \nu_{i-1}(\alpha) \sum_{\delta=-w}^{w} p_{\alpha}(\delta) \log_2 p_{\alpha}(\delta)\,.
\nonumber\\
\end{eqnarray}
Which in this case is equivalent with
\begin{eqnarray}
H(\mathbf{X})= H'(\mathbf{X})=\lim_{N \rightarrow \infty} H (X_N | X_{N-1}, \ldots , X_1) \nonumber\\
= - \lim_{N \rightarrow \infty}  \sum_{\alpha \in \text{CS}} \nu_{N-1}(\alpha) \sum_{\delta=-w}^{w} p_{\alpha}(\delta) \log_2 p_{\alpha}(\delta) \nonumber\\
= \sum_{\alpha  \in \text{CS}} \mu(\alpha) \cdot H(p_{\alpha} (\delta))\,,
\nonumber\\
\label{erate_qw_final}
\end{eqnarray}
where $\mu(\alpha) = \lim_{N \rightarrow \infty} \nu_N( \alpha) $ is the asymptotic distribution of coin states. Note that since in this paper we use the Hadamard coin matrix (\ref{coin_hadamard}) the (asymptotic) coin states do not form a continuous set, thus the use of a discrete sum over all coins states ($\text{CS}$) is sufficient.

In summary, the method of calculating the entropy rate is the following: First, one should determine the asymptotic coin distribution $\mu(\alpha)$. Then $p_\alpha(\delta)$ shift probabilities can be determined easily using formulas in Eqs.~(\ref{matrix_generalized}) and (\ref{pcdelta_def}). Finally, the entropy rate can be obtained using (\ref{erate_qw_final}). Note that the method proposed here can be applied directly for both finite or infinite systems. Also it can be extended in a straightforward way to higher dimensional quantum walks. However, such extension goes beyond the scope of the current paper.

We have not yet addressed the method for determining the asymptotic
coin distribution $\mu(\alpha)$. This can be done by defining a stochastic matrix,
\begin{equation}
P_{\alpha \rightarrow \beta} = \sum_{\delta=-w}^{w} \sum_{\chi \in \mathds{C}(\delta,\beta)} \delta_{\alpha,\chi} p_{\chi} (\delta)\,,
\label{coinstochmatrix}
\end{equation}
that is, the probability that from a $\alpha$ coin state after applying $U^w$ the walker is in the $\beta$ coin state after the position measurement.
It is straightforward to see that $P_{\alpha \rightarrow \beta}$ is indeed a stochastic matrix,
\begin{eqnarray}
\sum_{\beta \in \text{CS}} P_{\alpha \rightarrow \beta} & = & \sum_{\beta \in \text{CS}} \sum_{\delta=-w}^{w} \sum_{\chi \in \mathds{C}(\delta,\beta)} \delta_{\alpha,\chi} p_{\chi} (\delta)\nonumber\\
& = & \sum_{\delta=-w}^{w} p_{\alpha}(\delta) = 1\,.
\end{eqnarray}
After constructing the complete $P_{\alpha \rightarrow \beta}$ transition matrix, the $\mu(\alpha)$ asymptotic coin distribution can be readily found as the stationary state of the stochastic matrix $P_{\alpha \rightarrow \beta}$.

Moreover, 1D QWs have some symmetries which can be employed to make the calculation more efficient.
First, 1D QWs bear a spin-flip symmetry. This symmetry implies that,
compared to the general initial coin state $l | L \rangle_C + r | R
\rangle_C$, the orthogonal $r^{*} | L \rangle_C - l^{*} | R \rangle_C$
produces a mirrored position probability distribution. We use a single
important consequence of this property: A walk started from $| L
\rangle_C$ produces the exact same amount of entropy for any $w$ waiting times as
the walk started form $| R \rangle_C$, that is,
\begin{equation}
H\left( p_{L} (\delta) \right) =H\left( p_{R} (\delta) \right)\,.
\label{LRentropyequiv}
\end{equation}

Second, for 1D Hadamard QWs,
\begin{equation}
P_{\alpha \rightarrow L} + P_{\alpha \rightarrow R} \geq  2^{(1-w)} \quad\text{for all}\quad\alpha \in \text{CS}\,.
\end{equation}
Moreover, for arbitrary mixing coins of 1D QWs using the coin operator
\begin{equation}
C = \left(
\begin{array}{cc}
e & -f \\
f^{*} & e^{*}
\end{array}
\right)
\end{equation}
with $|f|^2 + |e|^2 = 1$ and $e,f \neq 0$
\begin{equation}
P_{\alpha \rightarrow LR} \equiv P_{\alpha \rightarrow L} + P_{\alpha \rightarrow R} \geq  |e|^{2(1-w)} \quad\text{for all}\quad\alpha \in \text{CS}\,.
\label{maptoLR}
\end{equation}
Here, we defined the summarized transition probability for the abstract ``joined'' coin state $LR$. This property has an immediate consequence: The black box based on a QW always forgets its initial state. Since from an arbitrary coin state a transition to $LR$ happens according to Eq. (\ref{maptoLR})
the part carrying information about the initial state $c_0$ at the iteration step $k$ is proportional to $(1-|e|^{2(1-w)})^k$, which is in the asymptotic $k \rightarrow \infty$ limit tends to $0$. This is one of our main results.

\begin{figure}[tb!]
\includegraphics[width=0.475\textwidth]{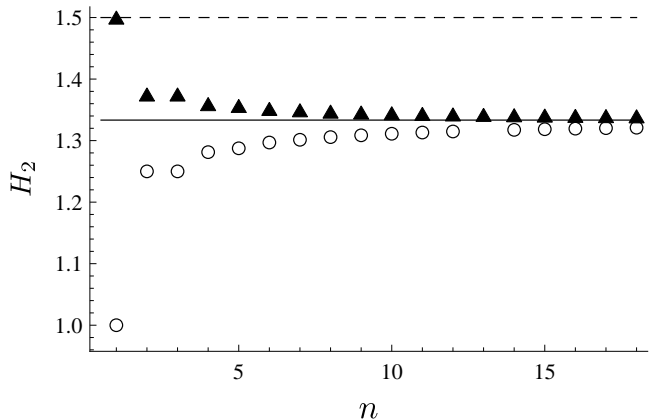}
\caption{
Convergence of the numerically calculated partial entropy rate  $H _2$ for $w=2$ waiting time.
We have evaluated the  definition of Eq. (\ref{eq:entrate}) for the first $n$ iteration steps, using the joint probability distribution
in Eq.~(\ref{joint_distrib}).
We used the 1D QW with Hadamard coin (see Eq. (\ref{coin_hadamard}) ); the triangles and circles correspond to the walk started from initial states $|\psi_0 \rangle = | 0, L \rangle$ and $|\psi_0 \rangle = \frac{1}{\sqrt{2}} \left( | 0, L \rangle +  | 0, R \rangle \right)$, respectively. The continuous line corresponds to the analytically determined rate for the simulated model: $H_2^{\text{QW}}=4/3$ bits. The dashed line corresponds to the rate of the CW : $H_2^{\text{CW}}=  3/2$ bits.
}
\label{Figure1}
\end{figure}

Using the method we gave above it is straightforward to determine the entropy rate of the QW with $w=2$ as the simplest, nontrivial case,
\begin{equation}
H^{\text{QW}}_2= \frac{4}{3}\,\text{bits.}
\end{equation}
The details of the exact calculation using this approach can be seen in Appendix~\ref{w2entrrate}.
For reference, the entropy rate of the CW for $w=2$ is $2/3$ bits as given by Eq. (\ref{entropy_classical}).
We numerically approximated the entropy rate using Eq. (\ref{eq:entrate}), for finite $n$'s (iterations). We found, that the obtained numerical data are converging to the rate we determined as illustrated in Fig \ref{Figure1}.
This result somehow contradicts the assumption that QWs generate more entropy because of the faster spreading. In fact, revealing the coin as a carrier of information, thus extracting more information from simple position measurement outcomes, allows for a more efficient prediction of the next step, essentially leading us to a more efficient source coding method --- and a lower entropy rate. However, it should be noted that for higher $w$ waiting times the entropy rate of the faster spreading QW inevitably surpasses the CW --- the proof behind this result is discussed in the following section.

The above given process is adequate when $\mu(\alpha)$ is nonzero for only a finite number of $\alpha$ coin states, i.e., the number of coin states arising under the full time evolution is finite or, equivalently, the size of $P_{\alpha \rightarrow \beta}$ is finite. However, depending on the coin operator and the waiting time we choose, the $P_{\alpha \rightarrow \beta}$ matrix can grow to infinite size. This issue can be solved by introducing a truncated (finite) basis; however, this will cause an uncertainty in the result.
We introduce a set of unknown coin states denoted by $| ? \rangle_C$,
which we use when we do not wish to calculate the elements of
$P_{\alpha \rightarrow \beta}$ further. In other words, the abstract set ``$?$" collects the coin states which the system does not touch up to the iteration step $k$, i. e., 
\begin{equation}
? = \left\{ \alpha \in CS \, | \, v_{i}(\alpha) = 0 \,\text{for all}\, i \in \left[0, k \right] \right\}\,,
\label{qmarkdef}
\end{equation}
where 
 $v_{i}(\alpha)$ is the coin state distribution at the $i$th iteration step as given in Eq.~(\ref{coindistrib}).
We note that rule of Eq. (\ref{maptoLR}) applies to ``$?$" as well, and it can be employed to make the truncated $P_{\alpha \rightarrow \beta}$ matrix a proper stochastic matrix.

Since the value $H(p_{?}(\delta))$ is unknown, Eq. (\ref{erate_qw_final}) cannot be used, but it can be bounded by giving an upper bound,
\begin{equation}
H_{\text{max}} = \max_{\alpha \in \text{CS}} H(p_{\alpha}(\delta))
\label{maxentr}
\end{equation}
and a lower bound,
 \begin{equation}
H_{\text{min}} = \min_{\alpha \in \text{CS}} H(p_{\alpha}(\delta))\,.
\label{minentr}
\end{equation}
Considering this, the value of the exact entropy rate (\ref{erate_qw_final}) are in the interval
\begin{eqnarray}
H(\mathbf{X}) &=& \sum_{\alpha \not{\in} | ? \rangle} \mu(\alpha) \cdot H(p_{\alpha} (\delta)) \nonumber\\ && + \frac{\mu(?)}{2} \left( H_{max} + H_{min} \pm \left\{ H_{max} - H_{min} \right\} \right) \,. \nonumber\\
\label{erate_qw_final_bounded}
\end{eqnarray}
Here we note that we use the compact form with a $\pm$ sign to denote the interval where the exact entropy rate resides.

The now proposed truncating method can be applied to approximate the entropy rates for arbitrary $w$'s. However, with 
the increasing of $w$ the size of stochastic matrices grows rapidly,
\begin{equation}
\mathrm{dim} \left( P_{\alpha \rightarrow \beta} \right) \approx \frac{1}{w-2} \left[ \left( w-1 \right)^{k+1} -1 \right] +1\,,
\label{approxdimmatr}
\end{equation}
where $k$ is the number of iterations of the procedure we take during the calculation
of the matrix $\left( P_{\alpha \rightarrow \beta} \right)$ --- and is also in the definition (\ref{qmarkdef}).  Similarly, the scaling of $\mu(?)$ can be approximated as it is proportional to the relative error of the calculated entropy rate. After a
lengthy, but straightforward, calculation this turns out to be
\begin{equation}
\mu(?) \approx (1-|e|^{2(1-w)})^{k+1} \,,
\label{approxmuquest}
\end{equation}
where we used $|e|^{2(1-w)}$ from  Eq. (\ref{maptoLR}).
Despite the exponential scaling of the precision and the dimension
with respect to the number of the iterations $k$, we found that our
method converge much faster than mere brute force simulations
reconstructing the joint probability distribution in Eq.~(\ref{joint_distrib}). This is due to the fact that the approximations (\ref{approxdimmatr}) and (\ref{approxmuquest}) are based on a worst-case scenario, while as it can be seen in the explicit calculations of this paper, the convergence of the $\mu(\alpha)$ distribution is much better.
To achieve an even better convergence, one can extend the proposed simplifications --- by use of the spin flip symmetry --- in order to find further isentropic states like the ones in Eq. (\ref{LRentropyequiv}).

\begin{figure}[tb!]
\includegraphics[width=0.45\textwidth]{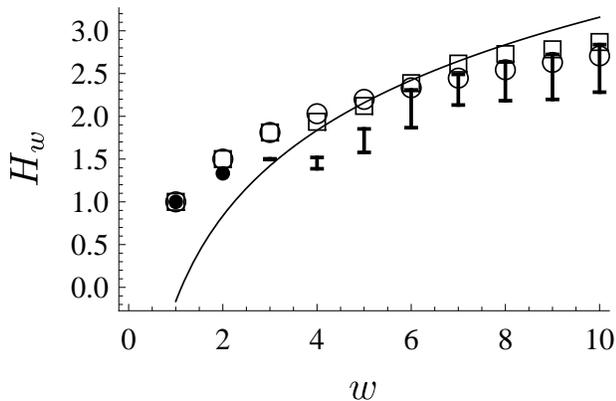}
\caption{ Entropy rates $H_w$ of the frequently measured walks on a 1D
  line as functions of waiting time $w$.  The circles correspond to
  the entropy rate $H_{w}^{\text{CW}}$ (see
  Eq.~(\ref{entropy_classical}) ) of the classical walk.  We used the
  Hadamard coin of Eq.~(\ref{coin_hadamard}) for the quantum walk.
  The black disks corresponds to the exactly calculated entropy rate
  $H_w^{\text{QW}}$ given by Eq.~(\ref{erate_qw_final}), while the
  vertical line segments correspond to the interval defined by the
  lower and upper bound on the entropy rate in Eq.~(\ref{erate_qw_final_bounded}).
 The rectangles correspond to the upper bound entropy rate
  $H_w^{\text{bound}}$ defined in Sec. \ref{naivesection}, while
  the continuous line represents the analytic approximation
  $H_w^{\text{approx}}$ of Eq~(\ref{erate_mixed_coin_approx}).
}
\label{Figure2}
\end{figure}

We determined the entropy rate of $w=3$ walks using the given methods. For the 1D Hadamard QW
the approximative calculation gave
\begin{equation}
H^{\text{QW}}_3 = 1.499 \pm 0.004 \approx 3/2 \,\text{bits}\,.
\end{equation}
In comparison, the CW walk has the entropy
rate of $H^{\text{CW}}_{3} = 3 - (3 \log_2 6) / (8 \log_2 2) \approx
2.031$ bits.
More details of the calculation based on the approximative method can be seen in Appendix~\ref{approxentrrate}.
We illustrate the results in Fig. \ref{Figure2}.

In the following we give an upper bound for the just now determined entropy rate which is easier to measure or compute. We will also discuss the scaling of the entropy rate of QWs with respect to waiting time $w$.

\section{Upper bound for the entropy rate of the 1D QW}
\label{naivesection}
Here we describe a method which will give us an easy-to-understand
and compute upper bound to the entropy rates of QWs.  If one is not
aware of the quantum nature of the walk  on which the information source (the
``black box'') is based, she or he might follow a measurement protocol
which is suitable for classical walks, thus ignoring the internal
quantum coin state. Such an absence of the information carried by the
coin leads to a less efficient encoding and, thus, higher entropy
rates. This statement is also supported by the fact that a function
of a Markov chain --- a hidden Markov chain --- has a higher or equal
entropy rate than the original chain \cite{Cover2006}, meaning essentially
an upper bound (and lower encoding efficiency).

Let us propose the measurement protocol which ignores the hidden coin
(memory) of the QW in the black box. Written in a standardized
manner, the protocol consists of the following steps:
\begin{enumerate}
\item Initialize the walker at state $| 0, c_{0} \rangle$, set position indicator $x = 0$.
\item Let the walk evolve for $w$ steps.
\item Perform a position von-Neumann measurement, which results a random position outcome $y \in [x \ominus w, x \oplus w]$.
\item Make a note that a $x \rightarrow y$ transition happened.
\item Repeat steps 2. - 5. with $y$ as the new $x$.
\end{enumerate}
Applying the protocol above for infinitely many times, the probabilities of  $x \rightarrow y$ transitions are calculated as relative frequencies. In this way, a stochastic transition matrix $P_{x \rightarrow y}$ describing the QW-driven process is obtained.
Note that in this way it is implicitly assumed that the system can be described via a time stationary classical Markov chain --- which is not true in general. Finally, the entropy rate is calculated using Eq. (\ref{entropy_stationary_markov}). 

One should note that for an infinite system the matrix $P_{x \rightarrow y}$ is not easy to handle.  However, QWs under consideration are translationally invariant ( cf. Eq.~(\ref{homogeneousity}) ).
Consequently,
\begin{equation}
 P_{x \rightarrow y} = P_{x+\delta \rightarrow y+\delta}
\end{equation}
for all $\delta$'s. Like in the classical case, we introduce $p(\delta)$ by Eq. (\ref{delta_equivalence}). Thus, the upper bound entropy rate $H^{\text{bound}}_w$ can be readily determined by Eq.~\eqref{delta_erate}, it is the Shannon entropy of the distribution of the arising position differences (shifts) in the stationary case.

We numerically calculated the upper bound for 1D QW and the actual entropy rate of 1D CW-driven black boxes using the above protocol.
 We used the Monte Carlo method to simulate the behavior of the black boxes, repeating the protocol until $p(\delta)$ appeared to converge. We found that $p(\delta)$ corresponding to the 1D QW in all cases converges to 
\begin{equation}
p(\delta) = \sum_{c={\{L,R\}}} \mathrm{Tr} \left\{  | \delta, c \rangle\langle \delta, c |U^w \tilde{\rho}_0 (U^w)^{\dagger} \right\} \,,
\label{completelymixing}
\end{equation}
where
\begin{equation}
 \tilde{\rho}_0 = \frac{1}{2} \left( \sum_{c'={\{L,R\}}} | 0, c' \rangle\langle 0, c' | \right) \,.
 \label{rhotilde0}
\end{equation}
Note that since $\rho_0$ is completely mixed in coin space, the effect of the initial $|c_{0}\rangle$ is lost, which is expected for a Markov chain. This result is in perfect agreement with our result given in the previous section: The system always forgets its initial state.

We found that for waiting times  $w \leq  3$ the upper bound rate of QWs with an unbiased coin coincides with the entropy rate of CWs, which can be viewed as classical correspondence in the strongly decohered limit. However, for the $w>3$ regime the upper bound surpasses the CW entropy rate, which is a direct consequence of the ballistic spreading.

To showcase the possible effects appearing on finite systems, we also
performed simulations  on finite $M$-cycles (1D cycle graphs with $M$
vertices). Increasing $w$ beyond $M/2$ in such a system causes an
interesting effect: The CW starts to evolve towards the uniform
distribution. As a consequence, the entropy rate becomes close to its
absolute bound $H_{\textrm{limit}}$ defined in
Eq. (\ref{entropy_limit}). In contrast to that, QWs  do not mix due to the unitary nature of the system.
Consequently, the self-overlap of the wave function might induce fluctuations in the entropy rate. In this self-overlapping regime the entropy production of CWs are usually higher.

\begin{figure}[tb!]
\includegraphics[width=0.475\textwidth]{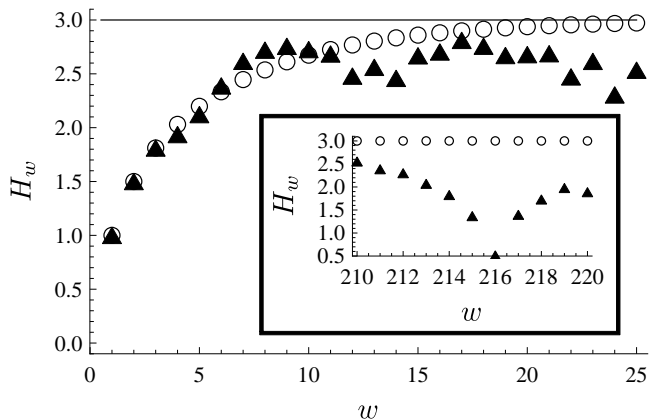}
\caption{
Entropy rate $H_w$ of periodically measured walks as the function of waiting time $w$.  
We used QW (triangles) with Hadamard coin ( see Eq.~(\ref{coin_hadamard}) ) and the unbiased CW (circles) on the cycle graph with 16 vertices. For the QWs we plotted the protocol giving the upper bound. The straight line corresponds to the theoretical entropy rate limit of Eq.~(\ref{entropy_limit}): $H_{\textrm{limit}} = \log_2 M - 1 = 3$ bits.
In the inset plot, we show traces of the collapse and revive like effects on the same system for high $w$ waiting times: For $w=216$ the time evolution operator comes very close to a simple permutation matrix, resulting in a very predictable behavior and an entropy rate upper bound $H^{\textrm{bound}}_{216} \approx 0.514$ bits. Meanwhile, the CW is totally mixing, resulting in an unpredictable outcome, with the maximal possible entropy rate $H_{\textrm{limit}} = 3$ bits.
We calculated all plotted data numerically using the Monte Carlo method until convergence occurred.
}
\label{Figure3}
\end{figure}

Increasing the $w$ waiting time even further, the unitary nature of
QWs eventually produces more interesting effects in finite systems: a
behavior similar to collapses and revivals \cite{Chandrashekar2010}
can be observed in the upper bound of entropy rate as a function of $w$ and in
the entropy rate itself. The appearance of these phenomena demonstrates the fundamental difference between the unitary and stochastic time evolutions in a black box.
 We illustrate these results in Fig.~\ref{Figure3}.

\begin{figure}[tb!]
\includegraphics[width=0.475\textwidth]{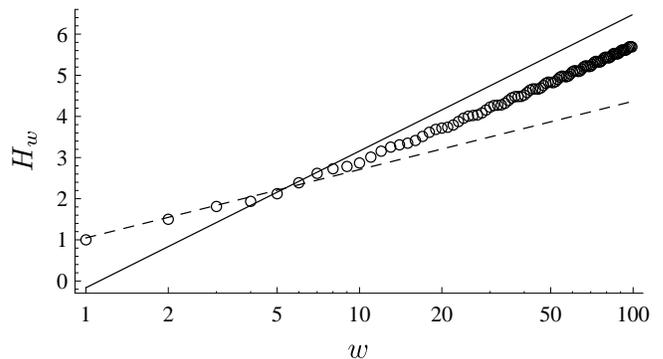}
\caption{
Upper bound of the entropy rate $H^{\text{bound}}_w$ of an 1D QW with Hadamard coin ( see Eq. (\ref{coin_hadamard}) ), denoted by circles, for infinite or finite ($w \ll M$) systems. We used high precision numerical simulations, and plotted the converged results. The dashed line corresponds to the analytically calculated entropy rate of CWs ( see Eq. (\ref{entropy_classical}) ), while the continuous line corresponds to the weak-limit-based approximation of Eq. (\ref{erate_mixed_coin_approx}). 
}
\label{Figure4}
\end{figure}

The observation in Eq. (\ref{completelymixing}) allows us to approximate the scaling of the entropy rate.
For the approximation we use the weak limit theory of quantum walks \cite{Watabe2008}. For high number of steps (high $w$'s) the symmetric probability distribution of a 1D Hadamard QW can be approximated with the formula
\begin{equation}
p(x,w) =\frac{1}{\pi  w \sqrt{1-\frac{2 x^2}{w^2}} \left(1-\frac{x^2}{w^2}\right)}\,,
\end{equation}
to be evaluated for $x \in [ -w/\sqrt{2}; w/\sqrt{2} ]$.
Note that this distribution corresponds to the walk started from the initial state localized at the origin, with a totally mixed initial coin state $\tilde{\rho_0}$ of Eq. (\ref{rhotilde0}), thus
\begin{equation}
p(\delta) = p(x,w) |_{x=\delta}
\end{equation}
Employing (\ref{delta_erate}) the upper bound of the entropy rate can be readily approximated by the integral
\begin{equation}
H_w^{\text{approx}} = - \int_{-w/\sqrt{2}}^{w/\sqrt{2}} p(x,w) \cdot \log_2 p(x,w) dx\,,
\label{approx_integral}
\end{equation}
which evaluates to
\begin{equation}
H_w^{\text{approx}} \approx -0.163164 + \log_2 w 
\label{erate_mixed_coin_approx}
\end{equation}
with high numerical precision.
It is apparent that the scaling of the upper bound of entropy rate goes with $\log_2 w$, in contrast with the scaling of the classical system ( cf. Eq. (\ref{entropy_classical}) ), which goes with $\log_2 \sqrt{w}$. This result can be interpreted as the consequence of the ballistic spreading of the QW. Our numerical test showed, that although the weak limit theorem predicts the $\log_2 w$ scaling, the scaling of the upper bound rate is for low waiting times are still close to $\log_2 \sqrt{w}$. Even for the regime around $w \approx 500$, we obtained scaling with
$\log_2 w^{0.94}$.
However, the scaling of the upper bound of the entropy rate for higher waiting times should converge to $\log_2 w$. We illustrate these results in Fig.~\ref{Figure4}.

Let us discuss the scaling of the exact entropy rate $H_w^{\text{QW}}$.
Using the weak limit approach calculating integrals similar to (\ref{approx_integral}) reveal the scaling for other initial states, i. e., the initial states giving the maximum and minimum entropy production $H_{\text{max}}$ and $H_{\text{min}}$ of (\ref{maxentr}) and
 (\ref{minentr}). In both cases the scaling is proportional to $\log_2 w$, thus  the precisely calculated $H_w^{\text{QW}}$ entropy rate is also scales with  $\log_2 w$ for high $w$ values.
 
In summary, the proposed protocol gives a straightforward way to measure, calculate, and approximate the upper bound of entropy rates of QW driven message sources. Since, the exact entropy rate can be quite hard to calculate, the easy-to-calculate and -measure upper bound is a proper tool for distinguishing walks with high waiting times $w$ living in a black box by their entropy production. 
We summarize the results given by all proposed methods in Fig. \ref{Figure2}.

\section{Analysis of independent systems --- the "most quantum" case}
When we gave the definition of the walker living in the black box, we explicitly stated that all measurements
are performed on the same system. However, for the case of QWs the frequent measurements mean loss of coherence, thus a step towards the classical world. One can easily create the ``most quantum" case, when at every iteration step the measurement is performed on a new, yet undisturbed system. Thus, in the first iteration step we perform a position measurement on a QW which took $w$ undisturbed steps, and then we discard the system. In the second iteration step, we perform a position measurement on another QW which took $2w$ undisturbed steps, and then we discard the system. All further steps are performed accordingly.
Thus, the $X_k$ sequence of stochastic variables is given by
\begin{equation}
p(X_k = x) = | S_x  U^{w \cdot k} | 0, c_0 \rangle |^2\,,
\end{equation} 
where $S_x$ is the position measurement projector given in (\ref{position_projector}) and $c_0$ is the initial coin state of the QW.
Since at every iteration step we perform measurement on a so-far undisturbed system, this is the ``most quantum'' case. However, this process erases all memory effects, and all correlations between subsequent measurements, thus all $X_k$'s are independent random variables.

Consequently, the entropy rate of such system is simply given by
\begin{equation}
H^{\text{mq}} = \lim_{k \rightarrow \infty} H(X_k)\,.
\end{equation}
Let us use our result about the scaling of the Shannon entropy of $H_{\text{min}}$ given in Sec. \ref{precapproachsection}.
The scaling of $H(X_k)$ is
\begin{equation}
H(X_k) \approx \log_2 k\,
\end{equation}
for 1D QWs in the infinite line.
Employing this, the entropy rate of the system is 
\begin{equation}
H^{\text{mq}} = \lim_{k \rightarrow \infty} H(X_k) = \lim_{k \rightarrow \infty}  \log_2 k = \infty \,.
\end{equation}
This is a straightforward consequence of the spreading of the system on an infinite line.
One can easily see that the entropy rate of both classical and quantum walks on infinite systems diverge to infinity, when we consider independent systems at each iteration steps.

However, one can address a question about the entropy rates on finite
systems. For the classical case on finite cycles with odd number of
edges, the entropy rate is given by Eq.~(\ref{entropy_limit}) due to
the mixing behavior of the system. However, since in the quantum case
the system is unitary, mixing does not occur but collapses and
revivals might appear as discussed in Sec.
\ref{naivesection}. Consequently, the entropy rate of independent
unitary QWs does not exist due to the lack of convergence. Similarly,
for 1D CWs on cycles with even number of sites, due to the oscillation
of the Shannon entropy limit given in Eq.~(\ref{entropy_limit}), the
entropy rate does not exist.  Summarizing the results, for the case of
the independent systems --- which is the most quantum scenario --- the
entropy rate is not a suitable tool for describing the asymptotic
information generation of the considered systems.

\section{Conclusions}

Entropy rate is an important quantity in classical information theory
which has a sound operational meaning: It is the asymptotic limit of
the lossless compression of the output of a discrete-time stochastic
process.  There can be many protocols in which some kind of dynamical
system produces a sequence of characters as output according to some
protocol, thereby realizing a classical stochastic process.  Here we
have studied an example and asked the question of whether the entropy
rate of a so-arising classical process captures some features of the
underlying dynamics (influenced, however, by the protocol).

The studied case involves a quantum walk, which is compared to one of
its classical limits. We have found that in this case the behavior of the entropy rate of the generated
classical stochastic process indeed differs for classical and
quantum walks and reflects some features of the underlying dynamics.

Although the classical definition of the entropy rate is extended to
quantum walks, the rich behavior of the quantum world is still
apparent.
We note that all results of the paper are given for 1D walks, but the
developed methods are more powerful and could be applied for more general systems.
We have given two approaches to calculate the entropy rates of such processes.
First, we described an elaborate method to
determine the exact entropy rates of 1D QWs. It turns out that in this
case the internal coin state --- which is not effected by the position
measurements --- serves as a memory, which allows us to develop a more
sophisticated coding, thus achieving a lower entropy rate.
In the case of frequent measurements the exactly calculated entropy
rate can be lower than the rate of classical walks, due to the
predictability provided by the coin state. 

Second, we gave an easy-to-measure and -calculate upper bound protocol that describes the entropy
production of 1D QWs when the observer is neglecting the quantum coin as
the memory of the walk.  In both cases the scaling of the entropy rate
tends to $\log_2 w$ for high $w$'s in contrast with the $\log_2 \sqrt{w}$
scaling of the classical walks, which is due to the ballistic spread
of the quantum model.  In both the exact and the upper bound
calculation we found that the entropy rate is independent of the
initial state of the 1D QW --- this is a particularly important result.

Both approaches can be employed to test the``quantumness" of the
frequently measured 1D walk residing in a black box; for low waiting
times ($w$'s) the exact entropy rate is easy to determine, thus it is
easy to distinguish between the classical and quantum models. For $w
\gg 1$ the $\log_2 w$ scaling of the rate corresponding to 1D  QWs can be
used as the indicator of quantumness. In this regime even the easy-to-calculate
upper bound measurement protocol should be enough to
distinguish between the classical and quantum walks.

We also investigated the``most quantum" scenario, when each
von Neumann measurement is performed on a new, undisturbed system,
thus, the system does not have memory. However, in this case the
entropy rate is not a suitable tool for describing per symbol
information generation due to the spreading nature of the system. Even
on finite systems QWs show no convergence due to the lack of mixing.

The fact, however, that the frequently measured 1D QW has a definite
entropy rate and can be described with a generalized stochastic matrix
also provides that the system which we have studied here can be
simulated with a well-designed classical walk, at least from the point
of view of the information content of the resulting classical
stochastic process. 
The question arises, and remains open for the time
being, if one can find dynamics and a protocol in which the behavior
or the mere existence of the classical entropy rate reflects some
fundamental nonclassicality.

\acknowledgements

We thank A. B. Frigyik for helpful discussions.
We acknowledge support by the Hungarian Scientific Research
Fund (OTKA) under Contract No. K83858 and
the Hungarian Academy of Sciences (Lend\"ulet Program,
LP2011-016). B. K. acknowledges support 
by the European Union and the State of Hungary, co financed by the
European Social Fund in the framework of T\'AMOP
4.2.4. A/2-11-1-2012-0001 ``National Excellence Program" (Nemzeti
Kiv\'al\'os\'ag Program). 
M. K. acknowledges the support of the grant
SROP-4.2.2/B-10/1-2010-0029 ``Supporting Scientific Training of
Talented Youth at the University of P\'ecs".

\appendix

\section{Entropy rate of 1D Hadamard walk for waiting time $w=2$}
\label{w2entrrate}
We show our calculation scheme for the entropy rate of QW driven ``black boxed" stochastic process $X_k$ ( Eq. (\ref{eq:entrate}) ), using the simplest nontrivial example of the 1D Hadamard walk, driven by the coin (\ref{coin_hadamard}).
Let us stick to the simplest case, when we initialized the walk in the coin state $| L \rangle_C$ at the origin, thus 
$c_0 = L$ and $w=2$.
We apply $U^2$ on $| \psi_0 \rangle = | 0, L \rangle $, resulting in the following
quantum state:
\begin{eqnarray}
U^2 | 0, L \rangle &  = & \frac{1}{2} | -2, L \rangle + \frac{1}{2} \left( - | 0, L \rangle + | 0, R \rangle  \right) \nonumber\\ && +  \frac{1}{2} | 2, R \rangle\,.
\end{eqnarray}
This yields some elements of $P_{\alpha \rightarrow \beta}$ ( see Eq. (\ref{coinstochmatrix}) ):
\begin{eqnarray}
 P_{L \rightarrow L} & = & 1/4 \nonumber\\
 P_{L \rightarrow -L+R} & = & 1/2 \nonumber\\
 P_{L \rightarrow R} & = & 1/4 \,.
\end{eqnarray}
We repeat this process again for the newly obtained coin states $| R \rangle_C$ and $\frac{1}{\sqrt{2}}(- | L \rangle_C + | R \rangle_C )$, thus we apply $U^2$ again, and we calculate new elements of the transition matrix $P_{\alpha \rightarrow \beta}$ as follows:
\begin{eqnarray}
 P_{-L+R \rightarrow L} & = & 1 \nonumber\\
 P_{R \rightarrow L} & = & 1/4 \nonumber\\
 P_{R \rightarrow L+R} & = & 1/2 \nonumber\\
 P_{R \rightarrow R} & = & 1/4 \,.
\end{eqnarray}
Note that in the second step of constructing the matrix, only a single new coin state $ \frac{1}{\sqrt{2}} (| L \rangle + | R \rangle)$ appeared.
Thus, we apply again $U^2$ on this new state to obtain the following:
\begin{eqnarray}
 P_{L+R \rightarrow R} & = & 1 \,.
\end{eqnarray}
We arrived to a complete coin state circle as no new coin states appeared, thus the $P_{\alpha \rightarrow \beta}$ transition matrix is complete. In the abstract coin state basis of $L, -L+R, R, L+R$ it takes the form
\begin{equation}
P_{\alpha \rightarrow \beta} = \frac{1}{4}\left(
\begin{array}{cccc}
1 & 2 & 1 & 0\\
4 & 0 & 0 & 0\\
1 & 0 & 1 & 2\\
0 & 0 & 4 & 0\\
\end{array}
\right)\,.
\end{equation}
$\mu(\alpha)$ is found readily as the left eigenvector of $P_{\alpha \rightarrow \beta} $ corresponding to eigenvalue $1$. Expanded in the same basis as the transition matrix, it takes the form of
\begin{equation}
\mu(\alpha) = \frac{1}{6}\left(2,1,2,1\right)\,.
\end{equation}
The single step missing is the calculation of the Shannon entropies $H(p_{\alpha}(\delta))$, which can be done in a straightforward manner, resulting in the following:
\begin{eqnarray}
H\left(p_{L}\left(\delta\right)\right) & = & \frac{3}{2}\, \text{bits} \nonumber\\
H\left(p_{-L+R}\left(\delta\right)\right) & = & 1  \,\text{bit}\nonumber\\
H\left(p_{R}\left(\delta\right)\right) & = & \frac{3}{2}\,  \text{bits} \nonumber\\
H\left(p_{L+R}\left(\delta\right)\right) & = & 1\,  \text{bit}\,.
\end{eqnarray}
Finally, employing Eq. (\ref{erate_qw_final}), the entropy rate is
\begin{equation}
H^{\text{QW}}_2 = \frac{4}{3} \, \text{bits}\,.
\end{equation}

In this particular example we restricted ourselves to initial state $| L \rangle_C$. One can repeat the process for a general initial coin state $| c_0 \rangle_C = l | L \rangle_C + r | R \rangle_C$ with $ | l |^2 + | r |^2 = 1$. After a more involving but still straightforward calculation it turns out that the size of $P_{\alpha \rightarrow \beta}$ is still finite in this case, and the entropy rate is $4/3$ bits independently from the initial coin state.
Moreover, this result holds true even for any mixed initial coin states.

We repeat the calculation of entropy
rate for $w=2$ from initial coin state $c_0=L = (LR)$ to demonstrate
the refined method using property (\ref{LRentropyequiv}). We write the
transitions corresponding to the abstract $LR$ coin state
\begin{eqnarray}
 P_{LR \rightarrow LR} & = & 1/2 \nonumber\\
 P_{LR \rightarrow -L+R} & = & 1/2 \,.
\end{eqnarray}
Investigating $\frac{1}{\sqrt{2}}(-| L \rangle_C+ | R \rangle_C)$ leads to
\begin{eqnarray}
 P_{-L+R \rightarrow LR} & = & 1  \,,
\end{eqnarray}
and, hence, we obtain the transition matrix
\begin{equation}
P_{\alpha \rightarrow \beta} = \frac{1}{2} \left(
\begin{array}{cc}
1 & 1 \\
2 & 0
\end{array}
\right)\,,
\end{equation}
in the basis of $LR$ and $-L+R$.  The asymptotic coin distribution $\mu(\alpha)$ turns out to be $\frac{1}{3}(2,1)$.
According to Eq.~(\ref{LRentropyequiv}), the Shannon entropy of $LR$ reads
\begin{equation}
H\left( p_{LR} (\delta) \right) =H\left( p_{L} (\delta) \right) =H\left( p_{R} (\delta) \right)\,.
\end{equation}
Thus, by employing Eq. (\ref{erate_qw_final}), we obtain
\begin{equation}
H^{\text{QW}}_2= \frac{4}{3}\,\text{bits}
\end{equation}
again.

\section{Approximating entropy rates of 1D QWs}
\label{approxentrrate}

In the following we
demonstrate the approximative method for determining $\mu(\alpha)$ for the case of $w=2$ and $c_0= L
(=LR)$. (Even though for $w=2$ the approximation is not necessary, it
is comparable with our previous consideration and it is easier to
follow than the $w>2$ cases.)  We restrict ourselves to the calculation of
the exact mapping only for the initial state, thus
\begin{eqnarray}
 P_{LR \rightarrow LR} & = & 1/2 \nonumber\\
 P_{LR \rightarrow -L+R} & = & 1/2 \,.
\end{eqnarray}
Since we do not wish to calculate further, using Eq.~(\ref{maptoLR}) we
get the following map
\begin{eqnarray}
 P_{-L+R \rightarrow LR} & = & 1/2 \nonumber\\
 P_{-L+R \rightarrow ?} & = & 1/2 \,,
\end{eqnarray}
where we used ``$?$" to mark the set of unknown coin states $| ?
\rangle_C$ which we do not wish to determine (see Eq. (\ref{qmarkdef}) ).  To build a proper
stochastic matrix we need an additional set of rules for the state
``$?$", which, using Eq.~(\ref{maptoLR}), are
\begin{eqnarray}
 P_{? \rightarrow LR} & = & 1/2 \nonumber\\
 P_{? \rightarrow ?} & = & 1/2 \,.
\end{eqnarray}
Thus, the transition matrix on the basis of $LR$ and $-L+R, ?$ is
\begin{equation}
P_{\alpha \rightarrow \beta} = \frac{1}{2}\left(
\begin{array}{ccc}
1 & 1 & 0\\
1 & 0 & 1\\
1 & 0 & 1
\end{array}
\right)\,.
\end{equation}
The corresponding asymptotic coin distribution is $\frac{1}{4}(2,1,1)$.
Using Eq. (\ref{erate_qw_final_bounded}) we finish our calculation,
in this particular case  $H_{\text{max}}=3/2$ bits and $H_{\text{min}}=1$ bit.
The exact entropy rate is in the interval
\begin{equation}
H_2^{\text{QW}}= 1.3125 \pm 0.0625 \,\text{bits}\,.
\end{equation}

We move to the case of $w=3$. For convenience, we use $c_0 = L (=LR)$ as the initial state.
We apply $U^3$ on $| \psi_0 \rangle = | 0, L \rangle$ resulting in the following:
\begin{eqnarray}
U^3 | \psi_0 \rangle & = & \frac{1}{\sqrt{8}}\left( |-3,L \rangle + | -1 \rangle_P \otimes \left( -2|L \rangle_C + |R \rangle_C \right) \right. \nonumber\\ && \left. - |1,L \rangle
 + |3,R \rangle  \right)\,.
\end{eqnarray}
Thus, we have transitions
\begin{eqnarray}
 P_{LR \rightarrow LR} & = & 3/8 \nonumber\\
 P_{LR \rightarrow -2L+R} & = & 5/8  \,.
\end{eqnarray}
Continuing with the new, yet undiscovered coin state, we obtain
\begin{eqnarray}
 P_{-2L+R \rightarrow LR} & = & 1/4 \nonumber\\
  P_{-2L+R \rightarrow 4L-3R} & = & 5/8 \nonumber\\
 P_{-2L+R \rightarrow L-2R} & = & 1/8  \,.
\end{eqnarray}
We end our calculation here and introduce the unknown coin state ``$?$" once again. Using Eq. (\ref{maptoLR})
we complete the transition matrix,  arriving at
\begin{equation}
P_{\alpha \rightarrow \beta} = \left(
\begin{array}{ccccc}
3/8 & 5/8 & 0 & 0 & 0 \\
1/4 & 0 & 5/8 & 1/8 & 0 \\
1/4 & 0 & 0 & 0 & 3/4 \\
1/4 & 0 & 0 & 0 & 3/4 \\
1/4 &  0 & 0 & 0 & 3/4 \\
\end{array}
\right)\,,
\end{equation}
which is written with respect to the basis  $(LR, -2L+R, 4L-3R, L-2R, ?)$.
From here, $\mu(\alpha)$, $p_{\alpha}(\delta)$ can be determined readily.
With the use of Eq. (\ref{erate_qw_final_bounded}),
the entropy rate for the 1D Hadamard QW is in the interval
\begin{equation}
H^{\text{QW}}_3 = 1.54 \pm 0.08\,\text{bits}\,.
\end{equation}

If we iterate the above procedure further, the interval (uncertainty) shrinks, i. e., the precision of the entropy rate increases. For $11$ iterations the entropy rate of the 1D Hadamard QWs with $w=3$ is
\begin{equation}
H^{\text{QW}}_3 = 1.499 \pm 0.004 \approx 3/2 \,\text{bits}\,.
\end{equation}

\end{document}